\documentclass{ws-procs9x6-cpt19}
\begin{document}

\newcommand{\refeq}[1]{(\ref{#1})}
\def\etal {{\it et al.}}

\title{Lorentz-Violating Running of Coupling Constants}

\author{A.R.\ Vieira$^1$ and N.\ Sherrill$^2$}

\address{$^1$Universidade Federal do Tri\^angulo Mineiro,
Iturama, MG 38280-000, Brazil}
\address{$^2$Department of Physics, Indiana University, Bloomington, IN 47405, USA}



\begin{abstract}
We compute the full vacuum polarization tensor 
in the minimal QED extension. 
We find that 
its low-energy limit is dominated by 
the radiatively induced Chern--Simons-like term 
and the high-energy limit is dominated by 
the $c$-type coefficients. 
We investigate the implications of the high-energy limit 
for the QED and QCD running couplings. 
In particular, 
the QCD running offers the possibility 
to study Lorentz-violating effects on the parton distribution functions 
and observables such as the hadronic $R$ ratio.
\end{abstract}

\bodymatter

\phantom{}\vskip10pt\noindent
Spacetime anisotropy affects not only clocks and rulers 
but also masses and couplings. 
Masses and couplings appearing in the tree-level Lagrangian 
are just parameters, 
which acquire corrections due to interactions. 
These parameters with their quantum corrections are 
referred to as the physical masses and couplings. Since quantum corrections are modified in the presence of Lorentz-violating effects, it is possible to place limits on Lorentz violation by studying the running of these quantities.

In addition, the coefficient space of
the QCD sector of the Standard Model Extension (SME) 
is comparatively unexplored.\cite{datatables} 
Therefore, 
it is of interest to study how 
Lorentz violation affects perturbative QCD processes like $e^+e^- \rightarrow \text{hadrons}$, deep inelastic scattering,\cite{DIS} the Drell--Yan process, and related quantities
like parton distribution functions (PDFs).\cite{Factorization}
In the following discussion,
let us consider the modified Lagrangian of a single-flavor fermion:
\begin{align}
\mathcal{L} =  &\tfrac{1}{2}i\bar{\psi}\Gamma^{\nu}\overset{\text{\tiny$\leftrightarrow$}}D_{\nu}\psi - \bar{\psi}M\psi -\tfrac{1}{4}F_{\mu\nu}F^{\mu\nu} \nonumber\\
& -\tfrac{1}{4}\left(k_{F}\right)_{\kappa\lambda\mu\nu}F^{\kappa\lambda}F^{\mu\nu} + \tfrac{1}{2}\left(k_{AF}\right)^{\kappa}\epsilon_{\kappa\lambda\mu\nu}A^{\lambda}F^{\mu\nu},
\label{EQ1}
\end{align}
where $D_{\mu}\equiv \partial_{\mu}+ieA_{\mu}$ is the usual covariant derivative, which couples the gauge field with matter,
\begin{align}
&\Gamma^{\nu}=\gamma^{\nu}+c^{\mu\nu}\gamma_{\mu}+d^{\mu\nu}\gamma_5\gamma_{\mu}+e^{\nu}+i f^{\nu}\gamma_5+\tfrac{1}{2}g^{\lambda\mu\nu}
\sigma_{\lambda\mu},
\label{EQ2}
\end{align}
and
\begin{align}
&M=m+m_5 \gamma_5 +a_{\mu}\gamma^{\mu}+b_{\mu}\gamma_5\gamma^{\mu}+\tfrac{1}{2}H_{\mu\nu}\sigma^{\mu\nu}.
\label{EQ3}
\end{align}

Considering the full fermion propagator is a difficult task. 
Instead, we treat the coefficients for Lorentz violation perturbatively by keeping only first-order corrections, which is appropriate
given existing experimental results. 
There is a total of six one-loop diagrams that 
correct the gauge-boson propagator. 
Also, 
the regularization scheme to be applied is a subtle task. 
As we can see in Eq.\ \refeq{EQ1}, 
some terms contain dimension-specific objects, 
namely the Levi--Civita symbol and the $\gamma_5$ matrix. 
The inadequate choice of a regulator 
may cause spurious terms, 
especially if we are dealing with the finite part of the amplitudes. 
Therefore, 
we apply a scheme called implicit regularization.\cite{Vieira}
This scheme allows us to stay in four dimensions 
and does not involve spurious symmetry-breaking terms 
in the process of renormalization. 
 
After computing the diagrams, 
we find that the low-energy limit ($p^2 \ll m^2$) 
of the renormalized vacuum polarization tensor 
is dominated by the induced Chern--Simons-like term
\begin{equation}
\Pi^{\mu\nu}_{\rm LV}(p)\approx \frac{e^2}{2\pi^2}\epsilon^{\alpha \beta\mu\nu}b_{\alpha}p_{\beta}.
\label{pilow}
\end{equation}
In computing Eq.\ \refeq{pilow}, 
we also find contributions from the
the $c$-type and $g$-type coefficients from Eq.~\eqref{EQ2}. 
However, 
they are suppressed relative to the dominant term by powers of $p/m$. 
Also, 
Eq.\ \refeq{pilow} can be affected by 
arbitrary and regularization-dependent surface terms. 
They are null if we require gauge invariance 
or momentum-routing invariance of the diagrams.\cite{Vieira} 

By contrast, in the high-energy limit $(p^2 \gg m^2)$ we find
\begin{align}
\Pi^{\mu \nu}_{\rm LV}(p)&\approx \frac{ie^2}{12\pi^2}(p^2 c^{\mu\nu}-p^{\mu}(c^{\nu p}+c^{p\nu})+(\mu \leftrightarrow \nu ))
\left( \ln \frac{-p^2}{m^2}-\frac{5}{3} \right)+ \nonumber\\
&+ \frac{ie^2}{6\pi^2} \left( \ln \frac{-p^2}{m^2}-\frac{13}{6}\right)c^{pp}\eta^{\mu\nu}+\frac{ie^2}{4\pi^2} c^{pp}\left(\eta^{\mu\nu}-
\frac{2}{3}\frac{p^{\mu}p^{\nu}}{p^2}\right),
\label{pihigh}
\end{align}
where $c^{p\nu}\equiv c^{\mu\nu}p_{\mu}$.
Equation \refeq{pihigh} shows that at high energies 
the running couplings will effectively
depend only on the $c$-type coefficients. 
It is also straightforward to show that
the Ward identity is fulfilled, 
$p_{\mu}\Pi^{\mu\nu}_{\rm LV}(p)=0$. 
If we insert this finite correction into a process, 
like electron--muon scattering, 
we can clearly see the $c$-type coefficients
affect the running of the QED coupling. 
The amplitude of this process takes the form 
\begin{align}
\mathcal{M}=-\frac{1}{p^2}[\bar{u}(p_3)\Gamma_{\mu}(p^2)u(p_1)]e^2_R(p^2)[\bar{u}(p_4)\Gamma^{\mu}(p^2)u(p_2)],
\label{amp2}
\end{align}
where $p^2 = (p_1 - p_3)^2$ and the renormalized charge is given by $e^2_R(p^2)=e^2_R(0)\left\{ 1+\frac{e^2_R(0)}{12\pi^2}\left[\ln \frac{-p^2}{m^2}-\frac{5}{3}
-\left(\frac{2c^{pp}}{p^2}\right)\left(\ln \frac{-p^2}{m^2}-\frac{2}{3}\right)\right]\right\}$. There is also an $e^2$ correction in the vertex,
$\Gamma_{\mu}(p^2)=\gamma_{\mu}+c_{\alpha \mu}\gamma^{\alpha}-\frac{e^2}{12\pi^2}c_{\alpha \mu}\gamma^{\alpha}
\left(\ln \frac{-p^2}{m^2}-\frac{5}{3}\right)$.

\def\sbs{\hsize=.45\textwidth\parindent=0pt\centering}
\long\def\sidebyside#1#2{\figurecaptionfont
\hbox to\textwidth{\vtop{\sbs#1\vskip1sp}\hfill\vtop{\sbs#2}}}

\begin{figure}[t]
\sidebyside
{
\includegraphics[width=1.65in]{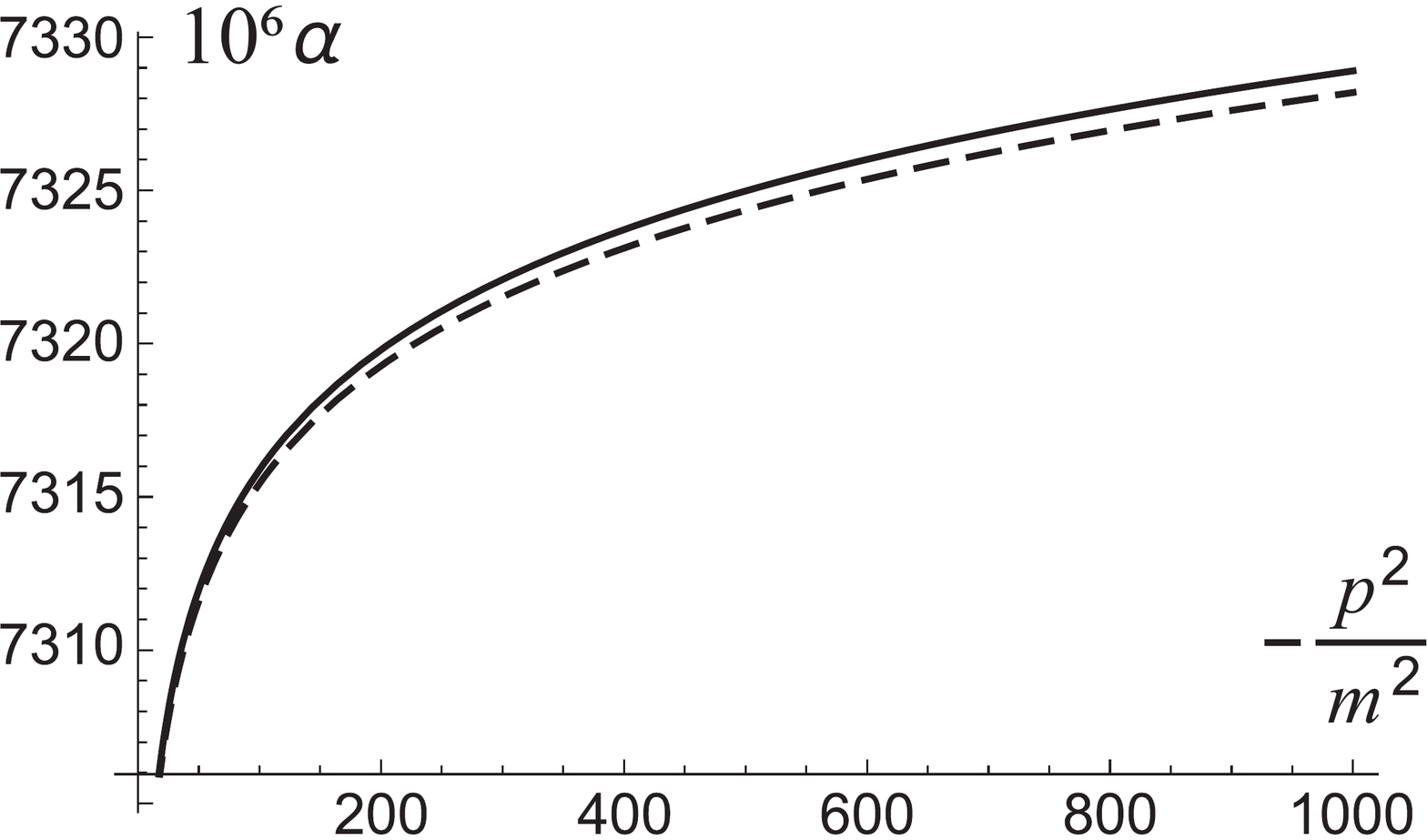}
\vspace{-1.6cm}
\caption{Running of the QED coupling. The dashed line is the running shifted by Lorentz violation for $\frac{c^{pp}}{p^2}\sim 10^{-2}$.}
\label{fig2.1}}
{
\includegraphics[width=1.65in]{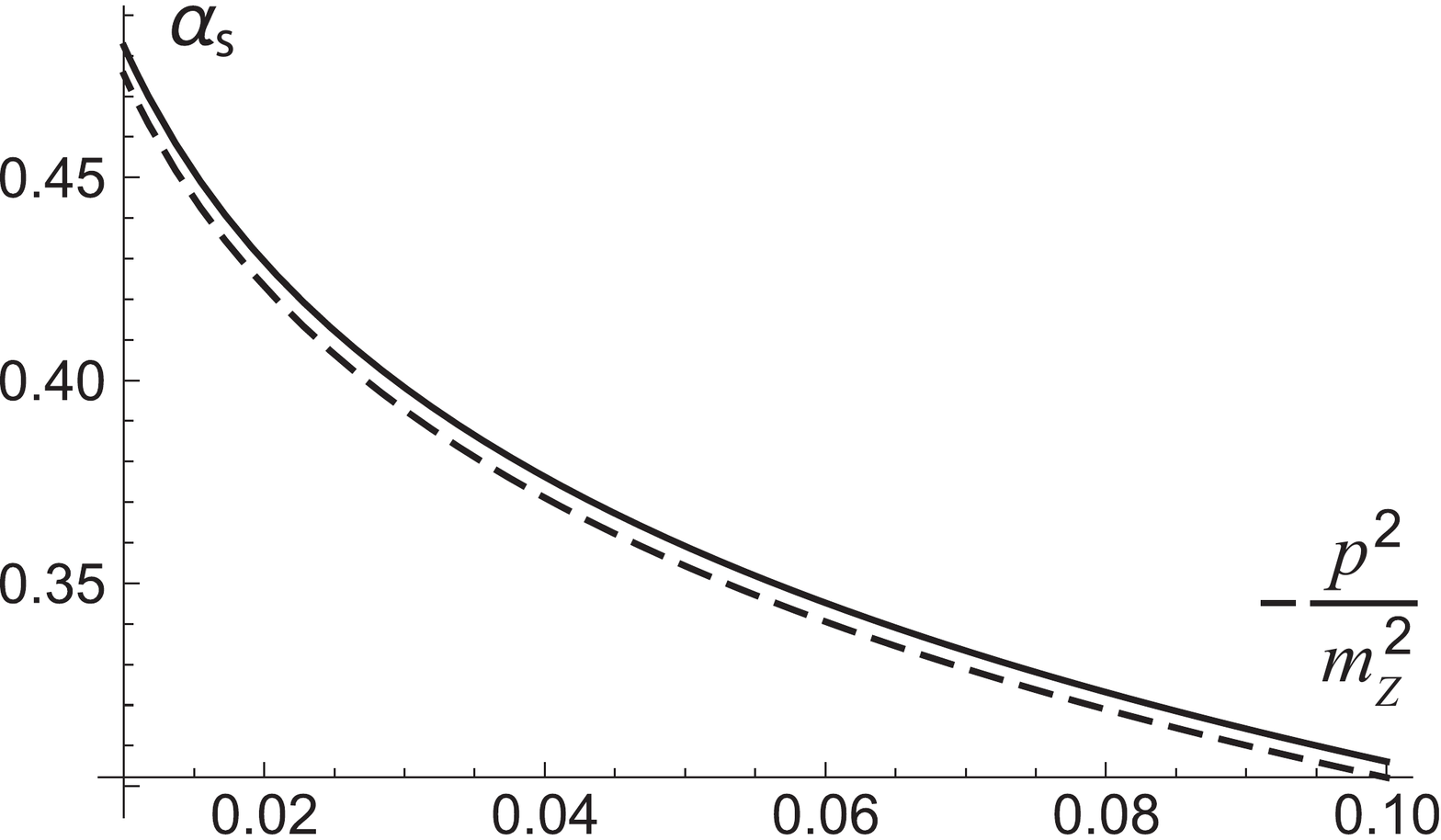}
\vspace{-1.6cm}
\caption{Running of the QCD coupling. The dashed line is the running shifted by Lorentz violation for $\frac{c^{pp}}{p^2}\sim 10^{-1}$.}
\label{fig2.2}}
\end{figure}

We run a simulation 
in order to see how the couplings evolve with these corrections. This is depicted in 
Figs.\ \ref{fig2.1} and \ref{fig2.2} 
for the QED and QCD coupling,
respectively. 
Bounds on the coefficients for Lorentz violation that 
come from $\alpha$ measurements are not so stringent; although $\alpha$ is very accurately known at the zero point, 
its measurements at higher energies 
have only a precision of two significant figures. 
However, 
we know the same thing happens in the running of the strong coupling $\alpha_\text{s}$
and QCD observables are affected by this running, 
so we can see how coefficients change these observables. 

The result of the high-energy limit in Eq.\ \refeq{pihigh} 
can also be used for quarks 
except for color and flavor factors. 
In the QCD computation, 
we also have to consider self-interacting gluon diagrams 
and the coefficients $k_G^{\mu\nu\alpha\beta}$ and $k_{AG}^{\mu}$. 
The former can be taken to be traceless, 
i.e., 
$k_{G\ \ \nu}^{\mu\nu\alpha}=0$, 
because a suitable choice of coordinates can absorb this term 
into $c^{\mu\alpha}$. 
The latter decreases with the energy scale 
so its effects are expected to be negligible at high energies.\cite{Colladay} 
In this way, 
we can focus on the $c$-type coefficients.

The Altarelli--Parisi equations 
depend on the running of $\alpha_\text{s}$. 
Hence, a perturbative correction of the type $c^{pp}/p^2$ 
implies the PDFs depend on Lorentz violation. It is interesting to compare
this observation with recent work that shows the leading-twist unpolarized PDF may implicitly 
depend $c^{pp}/\Lambda_{\text{QCD}}^2$, but instead through nonperturbative effects.
Quantum corrections also affect observables such as the hadronic $R$ ratio, since it
also depends on the running of $\alpha_\text{s}$. Here we find
\begin{equation}
  R= R_0\left\{ 1+\frac{\alpha_s (p^2)}{\pi}+\frac{\alpha^2_s(0)}{\pi}\frac
{2c^{pp}}{p^2}n_f \left(\ln \frac{-p^2}{\lambda^2}-\frac{2}{3}\right)+...  \right\},
\end{equation}
where $R_0$ is the tree-level ratio and $n_f$ is the number of flavors.

The QCD sector of the SME 
also affects the tree-level ratio $R_0$ 
besides the running of $\alpha_\text{s}$. 
There is a huge amount of data available on the $R$ ratio 
from the Particle Data Group. \cite{PDG} 
It is possible to run a simulation 
in order to constrain sidereal varations of coefficients that appear
using these measurements. 
We adapt the sidereal-time simulation of Refs.\ \refcite{DIS} 
for the first one hundred of these $R$ ratio measurements. 
Note that this sidereal simulation is only possible 
if we include quantum corrections 
because the tree-level Lorentz-violating correction 
is not energy-scale dependent 
as the $R$ ratio measurements are. 
We present 
the best limits in Table \ref{tab1}. 

\begin{table}
\tbl{Constraints on quark sidereal coefficients.}
{\begin{tabular}{@{}cc@{}}
   \toprule
    Coefficient & Constraint\\
   \colrule
    $|c^{XY}_q|$ & $<7.9\times 10^{-2}$ \\
    $|c^{YZ}_q|$ & $<3.4\times 10^{-2}$ \\
    $|c^{XZ}_q|$ & $<3.5\times 10^{-2}$ \\
    $|c^{XX}_q-c^{YY}_q|$ & $<1.6\times 10^{-1}$ \\
   \botrule
  \end{tabular}}
 \label{tab1} 
\end{table}

\end{document}